\documentclass[12pt]{article}
\begin{document}

\begin{titlepage}
\begin{center}
{\Large \bf Geometry of Time, Axiom of Choice and Neuro-Biological Quantum Zeno Effect}

\vspace{5mm}

\end{center}

\vspace{5 mm}

\begin{center}
{\bf Moninder Singh Modgil \footnote[1]{ Department of Physics,
Indian Institute of Technology, Kanpur, India.} }

\vspace{3mm}

\end{center}

\vspace{1cm}

\begin{center}
{\bf Abstract}
\end{center}
Role of axiom of choice in quantum measurement is highlighted by suggesting that the conscious observer chooses the outcome from a mixed state. Further, in a periodically repeating universe, these outcomes must be pre-recorded within the  non-physical conscious  observers, which precludes free will. Free will however exists in a universe with open time, It is suggested that psychology's binding problem is connected with Cantor's original definition of set. Influence of consciousness on material outcome through quantum processes is discussed and interesting constraints derived. For example, it is predicted that quantum mechanical brain states should get frozen if monitored at  sufficiently small space-time intervals - a neuro-biological version  of the so called quantum zeno effect, which has been verified in domain  of micro-physics. Existence of a very small micro-mini-black-hole in brain is predicted as a space-time structural interface between consciousness and brain, whose vaporization explains mass-loss reported in weighing experiments, conducting during the moments of death.

\vspace{1cm}

\end{titlepage}

\section{ Introduction}

	Consciousness is probably the most difficult problem attempted by human scientific endeavor, and is developing  into an eclectic discipline. In this paper, I introduce  certain new set theoretic ideas (among others) in the already inter-disciplinary field of  consciousness studies. At the outset, I shall clearly state relevant points  of my philosophical stance.
	\\
\noindent {\bf A. Nature of Consciousness.}

Conscious  observers are different from brain and bodies, but interact through them. This is identical to the dualistic school of thought, with Ecclles \cite {Ecclles 1977} as a representative. This apriori does not rule  out emergence of temporary consciousness in matter, as a result of various postulated mechanisms \cite{Crick 1994}, such as Bose-Einstein condensations, or self organizing behavior, or phase  locked dynamical  neural networks, strange attractors etc.. Whereas the material consciousness is temporary (depending upon stability of the physical  system), the non-material observer is eternal in time. This does not imply that observer is always conscious, or observer is conscious only in body. The term,  'Soul', is more accurate,  as consciousness and  observation are 'temporary phenomena' accompanying 'Soul', in certain conditions \footnote{For instancee, during dreamless sleep and coma, both consciousness and observations are absent}.  I shall however continue to use the term, "conscious observer", to represent soul.  Consciousness may thus be regarded as a common emergent property of both the  observer, as well as matter. As can be seen, the stance is flexible, enough to be compatible with  almost diverging views on the subject, and may  be termed as 'Non-dualistic'  in the sense that it allows simultaneous existence of almost divergent view points. Further, the conscious observers are distinct from each other, and inhabit the same (one) physical universe.

\noindent {\bf B. Nature of Time.}

	It is probably unlikely that complete problem of  consciousness can be solved without understanding nature of time. Time has been a subject of  many monographs and papers \cite{Zeh 1989, Sachs 1987, Mackey 1992} in  physics. The unresolved issues here are -
\begin{enumerate}
\item  Arrow of Time, i.e., its irreversibility,
\item Origin of present moment, alternatively the observed subjective distinction between, past, present and future, with conscious observers' attention confined to present (light-like hyper-surface), (ignoring out of body and similar psychic experiences \cite{Blackmore 1992} , for the time-being).
\item Overall geometry of time, i.e., cyclic or linear etc., \cite{Newton-Smith 1980}.
\end{enumerate}
	In periodic time, it can be shown that cause and  effect are connected by, what in set theory is called an equivalence relation \cite{Modgil 1995}. However, this causal structure could be important, as it has been  pointed out that  EPR paradox \cite{Einstein 1935}, which has been experimentally verified in recent years \cite{Freedman 1972, Clauser 1976, Fry 1976, Aspect 1981, Aspect 1982, Aspect 1984}, does suggest that causality is an equivalence relationship \cite{de Beauregard 1985}. While this circular geometry of time, would  appear to lead to the usual causal paradoxes of  time loops in  physics, such as  killing one's mother before one's birth (or conception!) - the causal paradoxes are avoided by withholding  free-will to the  conscious observers or actors or  participants (in the time cycle!). Thus, consciousness is only an observer of  events pre-recorded within itself, in an over all cyclic time, within this philosophical framework. Any apparent freewill is actually illusory.

	Further cyclic nature of time blurs the distinction between past and future, as the two are globally connected. In introduction of his monograph, Zeh \cite{Zeh 1989} quotes Lewis Carroll from the book "Through  the Looking Glass" -

White queen to Alice: \textit{"Its a bad memory which works only backwards"}.

In cyclic time, a good or perfect memory, will be able to  remember  all events which the possessor (conscious observer) would have observed with in the complete time cycle.

	I should add that while cyclic time has been  a  hallmark of ancient cosmological systems, Poincare , Zermelo, Caratheodoty,  and Nietzsche \cite{Poincare 1983, Zermelo 1919, Brush 1966, Nietzsche 1926} did attempt a mathematical  formulation of the  idea at the  turn of  the century, and the idea has  recently been evoked by contemporary mathematical physicists such as Segal \cite{Segal 1984}, Guillemin \cite{Guillemmin 1989}  and others \cite{Schchter 1977, Castell 1972} , with  a view  of solving certain problems of  observational cosmological physics, and particle physics (macro and micro cosmos). Idea behind introducing cyclic time concept is that if problem of consciousness is going to be solved [or the other way round), it may be so only in cyclic time - or what the  mathematical physicists would call the $S^1$ (circle) topology of time.

\section{ Set Theoretic Connection.}

	Cantor defined set as "collection  into a  whole, of objects  of our intuition  or  thought". The definition  is very psychological, and the phrase "collection into a whole" is of special relevance. The phrase actually implies  simultaneous perception of constituent set  elements as a whole. Even a sentence is understood only when perceived as a whole. Role of short term memory in verbal comprehension comes to mind when perceiving or comprehending very large sentences - here $7\pm2$ chunks of short term memory are  probably being  used  at  various levels, and the sentence has  to be read many times, before  comprehension (collection into a whole) occurs. Interestingly $7\pm 2$ has been derived by statistical mechanical  considerations, by developing so called Fokker-Plank equation  of brain neurons \cite{{Ingberg 1985}}. I equate verbal comprehension with (verbal) perception, as  such an equality is the reason for use of phrases such as "Now I see", when verbal comprehension occurs.

	While verbal  perception occurring in sentence comprehension involves only a small number of  elements, visual perception by contrast requires integration of a very large  number of  elements. Same can be said of other modes of perception, such as auditory, tactile, kinesthetic, olfactory, propeoceptory. Further, in the  conscious experience of an observer, these various perceptions from different senses, are further integrated into  a  whole, which even has non-sensory components such as  thoughts, memories, feelings, emotions. The complete momentary experience of a conscious observer is  thus  a set, in context of  Cantor's original definition. Further its a finite set, as it has a finite  number of elements, as represented by finite number of brain neurons.

\section{ Quantum Mechanical Aspects}

von Neumann \cite{von Neumann 1955}, Wigner \cite{Wigner 1966} and Pauli \cite{Pauli 1993}  have suggested the that wave-packet reduction was occurring  when the wave-packet was interacting with the conscious observer. Precise mechanism for this reduction by interaction  with consciousness has never been worked out. On the other  hand certain mechanisms for wave-packet reduction  have indeed been worked out by physicists, e.g., wave-packet reduction occurs when a coherent (unreduced) wave-packet approaches a system with infinite degrees of freedom \cite{Hepp 1972, Fakuda 1987}. Hawking \cite{Hawking} has also obtained interesting results concerning wave-packet reduction near a black-hole when a coherent state approaches it.

There exist other interpretations  of quantum mechanics  which try to do away with concept  of wave-packet reduction all together. These include Everett's many world interpretation \cite{Everett 1957} , Bohm's interpretation {Bohm 1957}, various hidden variable scenarios {Bellifante 1973} etc.. However, experimental evidence of quantum zeno effect {Itano 1990}, and  EPR suggests that wave-packet reduction is a good concept to explain  the results. I therefore assume that wave-packet reduction does occur, and  further it is caused by conscious observer. What are the properties of  consciousness which solves these quantum measurement problem? Let's refer to this set of  properties of consciousness as OM ( O - Observer, M - Measurement). Acronym OM has been selected, because of special significance word 'Om' has in context of search for one's true identity in ancient Indian philosophy.

von Neumann's model can be understood in terms of  "Quantum Measurement Chain" (QMC). In  this when an attempt is made to record state of say Schrodinger's cat, by a camera, the wave function of camera also passes into a coherent state. Same happens to the wave functions of film, human eyes, retina, brain neurons and so on. In von Neumann's interpretation, observer lies at the end of this quantum measurement chain, and leads to wave packet reduction of all the intermediary links (camera, brain, neurons) in this quantum measurement chain. This chain is  originating at Schrodinger's cat. Its possible for more than  one QMC to originate from the same system, e.g., when multiple observers are monitoring state of Schrodinger's cat. Such quantum measurement chaincs will be called linked.

Now, apriori there is no reason, why if  consciousness is causing wave-packet reduction (and influencing physical universe), it is not through the two mechanisms already outlined  by physicists Hepp, Fakuda, and Hawking (above). In absence of any other description of  process of consciousness causing wave-packet reduction, I accept what is not forbidden. Thus, as a conceptual agent responsible for wave-packet reduction, I attribute following  properties to consciousness - 

\noindent \textbf{OM-1. }Consciousness is a system of infinite degrees of freedom.

\noindent \textbf{OM-2. }Embodied Consciousness is associated with a black-hole.

Reason for use of letters \textbf{O} and \textbf{M} in \textbf{OM} will become clear in sections 4 and 5.

\section{ Black-hole in brain!}

	While popular notion  of black-holes as  massive astro-physical objects still awaits experimental confirmation, theoretical physics has moved ahead with concept of mini-black-holes, which are much less massive, and those which can be light enough to have mass of elementary particles, such as protons \cite{Recami 1992, Einstein 1935}. While the astro-physical black-holes are caused by gravitational collapse of stars whose  mass exceeds the so called Chandershekhar limit, the concept of micro-mini-black-hole (MMBH) is more like a singularity  or hole in fabric of space, i.e., its a place, where physical  space ceases to exist, so to say. Its gravitational influence is negligible, being  proportional to its mass, and so is its size. Its interesting because of its relativistic, quantum mechanical, and philosophical properties. The reader thus need  not be  alarmed, that all of his or her gray matter will be  sucked down this infernal black-hole in brain.

	There is  an interesting philosophical reason for existence of a black-hole associated with consciousness  in brain. While the soul or conscious  observer, is regarded as a  non-physical object, non-localizable in space-time, the present brain studies have almost localized it to be the region within brain. Black-hole provides a escape for  the non-material spirit. While the black-hole can be given physical  co-ordinates, the area within the event horizon, of the black-hole, effectively does not belong to the  physical universe - it lies  beyond the physical, universe so to say. Thus conscious observer located within such a black-hole strictly speaking does not exist  within physical space-time. In the event of physical death, the mind-body connection is severed, (e.g. Moody \cite{Moody 1977} , and one experts this black-hole to dissolve, or evaporate by a  mechanism, analogous to "Hawking radiation" \cite{Hawking 1974}. Its energy will be carried away by gravitational waves, and therefore will lead to a mass loss equal to mass of MMBH (few grams). Also effect of these gravitational waves generated at the moment of death, will be similar to high frequency acoustic waves, and would lead to cracking of any glass enclosure containing the physical body. Experiments verifying these phenomena have actually been done at the turn of this century \cite{MacDougall 1907}. The kind of tunnel vision reported in  near death experiences, i.e., motion through a long dark tunnel, with light  at the  end  of tunnel \cite{Moody 1977}, is actually in accordance with  optics near black-holes - an observer escaping  through a black-hole would  actually experience, similar tunnel vision!

	Black-holes have many other  interesting properties such as existence of local closed time like curves, {Morris 1988} which could explain ability  of clairvoyants to see past and future events, while existing  within the physical body (by motion of point like conscious observer, on one of the local closed time like curves near MMBH). Black holes also provide a handle, or gateway to other  dimension, and non-physical universes, which  are of special interest Brahma Kumaris and possibly other workers in Transcendental. Interestingly, tachyons (particles traveling faster than light) falling through a  black-hole, leave the conventional physical universe  of three space and one time dimension, and  enter  a  universe with three time and  one space  dimension {Chandola 1986}. This latter universe or rather meta-universe,  could  be of special interest for actual  meta-physical  experiences.

\section{ Records within Consciousness}

	Apriori, von Neumann's interpretation of consciousness causing wave-packet reduction, does not determine, as to which particular outcome is actually selected. Neither do any other mechanisms such as  Fakuda's, Hepp's or Hawking's - all they do is reduce a coherent superposition of states into an incoherent mixture - actual outcome then being a psychological process of observation. Thus if embodied consciousness is monitoring quantum states of 109  neurons, and causing their wave-packet reductions, which leads to perception, and the complete momentary experience of that observer, then these wave-packet reductions must be recorded within the conscious observer, in a consistent  fashion, and cannot be random because - random reductions, will not lead to the  perceived order  of the universe. Various laws of physics, such as those of continuity, conservation, invariances, etc., are result  of perceptions, based upon  these wave-packet reductions. Hence we  can identity another property of  consciousness -

\noindent \textbf{OM-3.} Recorded within conscious observer is outcome of all quantum measurements performed by it (as reflected in coherent  brain states). These reductions are not random, but have a logical relation to each other, which is the basis for invariances observed in physics, and results of EPR and Bell's inequality experiments etc.

	Now this selection of a particular  outcome, from the set of all possible outcomes is another psychological  process, which was encountered quite early ;in development of set theory. Its called Axiom of Choice \cite{Moore 1982}. Briefly, it states that, given a set, there exists a choice function, which selects an element  of the set. Only problem is, that while it is vital to almost all of mathematics, its use has lead to paradoxical Banach-Tarski theorems, involving duplication of spheres, and show that concept of additive measure is not sound \cite{Wagon 1985}. Details of how axiom of choice relates to quantum measurement process are worked out in  \cite{Modgil 1992}

	There exists another reason for records within consciousness, and evidence from it  comes from following scenario reminiscent of EPR of quantum mechanics. The argument has a philosophical flavor - if distinct observers A and B separated by a space-like interval, observe state of Schrodinger's cat in an  experiment at a  particular instant, it is required that both should observe it either dead  or alive. It should not be that observer A, finds cat alive, and observer B finds cat dead. Thus if  wave function of Schrodinger's cat as represented  in brain neurons of  observer  A is collapsing, due to a recording within A, this collapse is compatible with a similar collapse occurring in brain of  observer B. To ensure this compatibility we require -

\noindent \textbf{OM-4.} Recording within conscious observers with respect to measurement performed on the  same quantum system are mutually compatible. Alternatively, outcomes observed by observers lying at ends of distinct but linked  quantum measurement chains, are compatible.

	Support of EPR results of quantum mechanics for  records within  consciousness is as follows. Lets say that  observers A and B are separated by a space-like interval, and  perform measurement on a  correlated  photon pair. EPR results indicate that wave functions of both the photons are collapsing only at the  moment of measurement, and the  collapses are mutually compatible, which A and B will also notice, when they compare notes latter on. Now, there  exists  no way for observer A  to send a signal to B, regarding his or her outcome, within the  frame work of present day physics (light speed limit  and all that). The question  therefore exists, if the conscious observers  A and B are indeed causing these distinct but correlated collapses, how is mutual compatibility being ensured? OM3 is thus related to OM4. This is also where cyclic nature of time may be playing  an important role. In cyclic time, the same measurements would have been performed in all the past (infinite) time cycles, and identical outcomes would have been recorded, in all the time cycles. This geometry of time, appears to provide at least a chance for observers to compare note and  correlate their outcomes. See \cite{Modgil 1995} for  a possible scenario of  communication between observers in EPR type experiments using light like signals  in cyclic time.

	I close this section, with an argument, for why the embodied consciousness should exist in a MMBH (micro-mini-black-hole). If outcomes of quantum measurement are pre-recorded in the conscious observers, (OM-3 and OM-4) such a recording constitutes "hidden variable \cite{Bellifante 1973}  determining the quantum measurement outcome". Now Bell's theorem \cite{Bell 1964} yielded inequalities, which would distinguish between hidden variable scenario, and actual quantum mechanics (without hidden variables). Experiments \cite{Freedman 1972}-\cite{Aspect 1984}  to test between the two yielded results in accordance with quantum mechanics, i.e., no functions or additional physical  hidden variables, which would determine the  outcome. Locating the consciousness within a black-hole resolves this problem, because now the recording (hidden variable determining the outcome) is lying beyond the event  horizon of the black-hole, and thus effectively outside the universe, and therefore is beyond the purview of present formulation of  Bell's theorem \cite{Bell 1964}.

\section{ Neuro-Biological-Quantum-Zeno-Effect}

	Readers would be familiar with ancient Greek Zeno's paradox \cite{Mishra 1984}, which questioned the concept of motion, by arguing that if an arrow in flight was being continuously observed, and occupied a position at every time instant, as to how could  the apparent motion observed was actually possible? Though the paradox was resolved in continuum based classical mechanics, it  has reappeared in grab of quantum zeno effect (QZE) - so christened by physicist Sudarshan \cite{Mishra 1984}. Briefly, if a system is in state A, and about to change to state B, before it does so, its wave function (mathematical object describing its physical state), has to  go  into  a superposition of states A and B, i.e., the system exists in a sort of (A+B) state. Now when a quantum measurement is  performed on this superposed (A+B) state, the wave function collapses to either A or B. So, if the  measurements are being performed sufficiently rapidly, the wave function of the  system cannot evolve to first state (A+B) and than state B. As a result it remains frozen in  state A. This effect is also called "watch dog effect", \cite{Joos 1984} (thief moves only when watch dog closes its eyes), and the ''boiled kettle phenomena", (kettle appears to boil over and spill, just when one's attention is diverted). Thus in terms of quantum mechanics, paradox of  Zeno's arrow can be formulated and resolved as follows. Where as, wave-function of arrow, which also  describes its  position is evolving  continuously,  the actual act of wave-packet reduction, by monitoring  or perception of a conscious observer, is a non-continuos phenomena. This is  because the process of human perception requires  a large number  of photons  from Zeno's arrow to  reach human eye and retina, where after a time delay, a signal is relayed to brain, and a quantum mechanical  representation of  arrow's (superposed) wave function  is formed by observer's neurons. This quantum measurement chain of coherent (superposed) state collapses when the  conscious observer perceives arrow's position, and is  a  non-continuos phenomena. Thus, in between these non-continuos perceptions, wave function of arrow can evolve to different positions. QZE has been verified for ensembles of atoms about to  make electron transitions from a higher energy state to a lower energy state [36]. Monitoring at progressively  smaller intervals, reduces actual  number  of atoms making  the transition, in a given period of time.

	Neuro-biological-quantum-zeno-effect (NBQZE), as the term suggests, implies that, if  brain state of a  person is being  monitored at sufficiently small space-time scales, (by another person, with data being recorded onto say a computer, all of which is  later examined by the experimenter), then neurons of the subject will not be able to  evolve to a coherent state and make transition, to another  state which would represent transitions from one perception to another. Thus person's subjective experience would be blocked. Even though external sensory stimulus may be  applied, the subject would not report perception of the stimulus. The resultant state would be similar to highest states of meditation, which involve complete withdrawal of consciousness from the body and senses - effectively the  consciousness has ceased to interact with the physical universe, and is no longer performing any quantum  measurement - wave-packet reductions, in his or her brain are being caused  by the experimenter, and are preventing brain state from evolving along with the  wave function of the  changing environment or universe.

\end{document}